\documentclass[oupdraft]{bio}
\usepackage{url}

\usepackage{amsfonts}
\usepackage{amsmath}
\usepackage{bm}
\usepackage{booktabs}
\usepackage{placeins}
\usepackage[caption=false]{subfig}

\begin{document}

\title{A simulation study of disaggregation regression for spatial disease mapping}

\author{ROHAN ARAMBEPOLA$^\ast$, TIM C. D. LUCAS, ANITA K. NANDI\\[-6pt]
\textit{Big Data Institute, Nuffield Department of Medicine, University of Oxford, Oxford, UK}\\[-4pt]
{rohan.arambepola@stx.ox.ac.uk}\\[4pt]
PETER W. GETHING, EWAN CAMERON\\[4pt]
\textit{Telethon Kids Institute, Perth Children’s Hospital, Perth, Australia\\[3pt]
Curtin University, Perth, Australia\\[-8pt]
Big Data Institute, Nuffield Department of Medicine, University of Oxford, Oxford, UK}}




\markboth%
{R. Arambepola and others}
{A simulation study of disaggregation regression}

\maketitle

\footnotetext{To whom correspondence should be addressed.}

\begin{abstract}
{Disaggregation regression has become an important tool in spatial disease mapping for making fine-scale predictions of disease risk from aggregated response data. By including high resolution covariate information and modelling the data generating process on a fine scale, it is hoped that these models can accurately learn the relationships between covariates and response at a fine spatial scale. However, validating these high resolution predictions can be a challenge, as often there is no data observed at this spatial scale. In this study, disaggregation regression was performed on simulated data in various settings and the resulting fine-scale predictions are compared to the simulated ground truth. Performance was investigated with varying numbers of data points, sizes of aggregated areas and levels of model misspecification. The effectiveness of cross validation on the aggregate level as a measure of fine-scale predictive performance was also investigated. Predictive performance improved as the number of observations increased and as the size of the aggregated areas decreased. When the model was well-specified, fine-scale predictions were accurate even with small numbers of observations and large aggregated areas. Under model misspecification predictive performance was significantly worse for large aggregated areas but remained high when response data was aggregated over smaller regions. Cross-validation correlation on the aggregate level was a moderately good predictor of fine-scale predictive performance. While the simulations are unlikely to capture the nuances of real-life response data, this study gives insight into the effectiveness of disaggregation regression in different contexts.}
{Disaggregation; Downscaling; Disease mapping; Bayesian hierarchical modelling; Geostatistics.}
\end{abstract}

\section{Introduction}
\label{sec1}
High resolution maps of disease risk are an important public health tool, facilitating efficient allocation of limited resources and precision targeting of interventions \citep{drake2017geographic, elliot2000spatial, lawson1999disease}. Where incidence or prevalence data is available at a point-level (or can be treated as such, as often the case with village-level health surveys), traditional geostatistical models can be used to make fine-scale (or `pixel-level') predictions of risk over the region of interest (typically on a grid of 1km-by-1km or 5km-by-5km pixels), informed by spatial patterns and environmental and socioeconomic covariates \citep{diggle2003introduction}. Much of the information in these models comes from the spatial smoothing of the data, particularly near sampling locations (some of the earliest examples of modern spatial mapping used no covariate information, such as \cite{diggle1998model}). Further from these locations, spatial smoothing is less informative and covariate information aids more in predictive performance. Often, however, only aggregated (or `polygon-level') response data is available, for example case counts in geographical regions or from health facilities. Attempting to make statistical inferences at this aggregate level and subsequently predict at fine scale has two main pit-falls --- the `ecological fallacy' \citep{wakefield2006health}, where relationships learned between covariate and response variables at the aggregate level may not hold on a fine scale, and the `modifiable areal unit problem' \citep{fotheringham1991modifiable}, where fine-scale predictions based on aggregated data change based on how the data is aggregated geographically. 

Disaggregation regression \citep{keil2013downscaling, sturrock2014fine, weiss2019mapping, taylor2018continuous} attempts to avoid these problems by using high resolution covariate information and modelling the data generating process on a fine scale, with the likelihood of the observed data given by the sum of these processes. However, it is often difficult to assess the performance of these models due to a lack of observed data on a fine spatial scale with which to validate fine-scale predictions. For this reason it is unclear in which situations disaggregation regression may be successfully applied and when making accurate fine-scale predictions may not be possible. Providing fine-scale risk maps where there is insufficient data to inform these predictions (for example, where response data is aggregated over very large areas or information is only available from a small number of areas) may be misleading and in these cases aggregated risk maps may be more suitable. Cross validation can be done on the aggregate level (as by \cite{lucas2020mapping, law2018variational}, for example) but again it is not clear how measures of out-of-sample aggregated performance relate to fine-scale predictive accuracy. 

There have been a number of important simulation studies of disaggregation methodology, generally serving as proofs of concept of particular methods. \cite{wilson2020pointless} perform a simulation study using their disaggregation method (combining point and polygon-level data) with no covariates in a small number of spatial settings, while \cite{li2012log} compare disaggregation regression to regression on the aggregate level with two covariates (one categorical and one continuous) on a single set of areal units. \cite{law2018variational} simulate data using a toy dataset and evaluate performance at the aggregate level when using real world malaria data. This study builds on the existing work in two ways. Firstly, we extend the scope of previous simulation studies by evaluating the performance of disaggregation regression in a realistic setting and under different conditions to better inform the utility of these methods when applied to real world datasets. Each situation is repeated a number of times to ensure the robustness of the results. Secondly, we investigate the use of cross-validation on the aggregate level as a metric of fine-scale predictive performance. Cross-validation metrics on the aggregate level are often used due to lack of fine-scale observations, so it is important to understand how well these aggregate metrics represent fine-scale accuracy.

This study provides insight into the performance of disaggregation regression as the size of aggregated areas, number of observations and the level of model misspecification varies. In each case, incidence data was simulated for each pixel and aggregated to give polygon-level response data. Disaggregation regression was then used to make pixel-level predictions from this aggregate data, with the simulated fine-scale data used as a ground truth (which is typically missing in real world applications). Additionally, we calculated polygon-level correlation using $k$-fold cross validation and compared this measure to the fine-scale predictive performance. 

The simulated data was intended to resemble malaria incidence data, using real covariate information (a mix environmental and socioeconomic variables that are commonly used in malaria risk mapping) and geographical information from two malaria endemic countries, India and Madagascar. However, we believe the results of this study are informative for disease mapping in general and more widely. We considered model misspecification due to unobserved covariates but did not consider other possible sources of misspecification, such as non-linear covariate effects, covariate interaction or alternative case-generating processes.
\section{Methods}
\label{sec2}
\subsection{Data simulation}
The number of cases in pixel $j$ of polygon $i$ was drawn from a Poisson process with mean, $\mu_{ij} = \lambda_{ij}\times p_{ij}$, given by the product of the pixel incidence rate, $\lambda_{ij}$, and pixel population, $p_{ij}$. The log pixel incidence rate was simulated as a linear combination of observed, $\mathbf{X}_{ij}^\text{obs}$, and unobserved, $\mathbf{X}_{ij}^\text{unobs}$, covariates
$$\log\lambda_{ij} = \beta_0 + \bm{\beta}_\text{obs}^T\mathbf{X}_{ij}^\text{obs} + \bm{\beta}_\text{unobs}^T\mathbf{X}_{ij}^\text{unobs}.$$
The elements of $\bm{\beta}_\text{obs}$ and $\bm{\beta}_\text{unobs}$ were drawn independently from a Normal distribution with mean $0$ and standard deviation $0.5$. The value of the intercept $\beta_0$ was drawn uniformly between -8 and -5 in order to produce incidence rates similar to those observed in malaria endemic countries. The sampling process was repeated if the total case counts were above country-specific upper thresholds, to ensure that the total cases were largely consistent with estimated annual malaria case counts in each country \citep{world2019world}.

Mock case data were simulated under three different scenarios, representing different levels of model misspecification. In each scenario, the observed covariates were 6 real covariate surfaces. In scenario 1, there were no additional unobserved covariates and therefore no model misspecification. In scenario 2, there were 6 additional real covariates which were unobserved during the fitting process. Finally, in scenario 3 there were 6 additional real covariates and 3 mock covariates that were unobserved during the fitting process. See Section \ref{study_area_and_covariates} for more details on the real and simulated covariates used. For each country and scenario, 20 risk surfaces were simulated. For each simulated risk surface, the observed covariates were sampled (without replacement) from the 12 possible real covariates. In scenario 2 the unobserved covariates were the remaining real covariates while in scenario 3 the unobserved covariates were the remaining real covariates and 3 mock covariates, sampled without replacement from the 12 possible mock covariates for each risk surface.

Case numbers in each pixel were aggregated to give polygon-level case counts. These polygons were the real administrative units in each country and model performance was investigated using three levels of subdivision - administrative level 1, 2 and 3 units (shown in Figure \ref{fig:shapes}). Administrative level 1 units are the largest subnational administrative regions (for example, states and union territories in India) while administrative level 2 and 3 units are increasingly fine subdivisions. 

\subsection{Study area and covariates} \label{study_area_and_covariates}
\begin{figure}
\centering
  \includegraphics[width=0.75\linewidth]{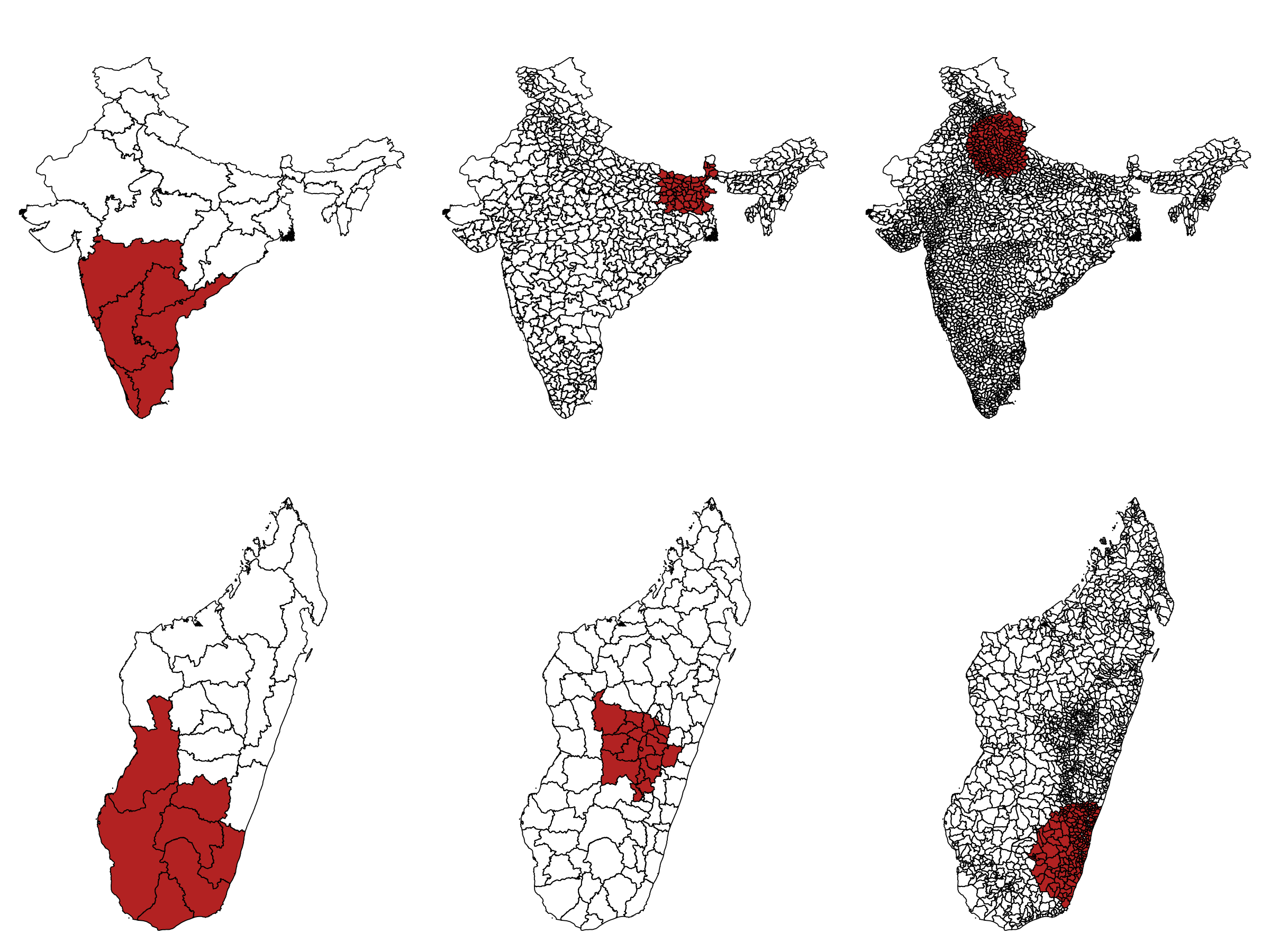}
  \caption{Administrative level 1, 2 and 3 units (left to right) in India (top) and Madagascar included in this study. Highlighted in red are examples of the randomly sampled contiguous sets of polygons which were used in Section \ref{exp1}.}
  \label{fig:shapes}
\end{figure}

The countries chosen for this study were India and Madagascar. These two settings provided a range of polygon sizes (at each administrative level regions were generally larger in India) and different environmental and socioeconomic profiles and therefore covariates with different spatial patterns. Islands were not included and two union territories in India (Chandigarh, and Dadra and Nagar Haveli and Daman and Diu) were excluded as these regions are much smaller than the typical administrative level 1 unit. See Figure \ref{fig:shapes} and Table \ref{admin_table} for more information on the administrative units used in this study.  Shapefiles for these administrative regions were obtained from the Malaria Atlas Project database using the malariaAtlas R package \citep{pfeffer2018malariaatlas}. 

\begin{table}
   \caption{Summary of sub-national administrative regions.\label{admin_table}}
 \centering
\begin{tabular}{@{}lllllll@{}}
\toprule
\multicolumn{1}{l}{} & \multicolumn{3}{l}{India} & \multicolumn{3}{l}{Madagascar} \\ 
Administrative level   & 1         & 2        & 3        & 1       & 2       & 3     
\\ \midrule
Number                 & 32         & 660        & 2,286        & 22       & 113       & 1,425      \\
Mean size (km$^2$)        & 98,000         & 4,800        & 1,400        & 27,000       & 5,200       & 400    
\end{tabular}
\end{table}

\begin{table}
    \caption{List of covariates.\label{cov_table}}
 \centering
\begin{tabular}{l p{0.36\textwidth} p{0.35\textwidth}} \toprule
Covariate & Description  & Source\\ \midrule
Accessibility & Distance to cities with population $>$ 50,000 & \cite{weiss2018global}\\
Aridity & Aridity index &  \cite{trabucco2009global}\\
Elevation & Elevation as measured by the Shuttle Radar Topography Mission & \cite{farr2007shuttle}\\
EVI & Enhanced vegetation index &   \cite{modisEVI} \\
Friction &  Generalized rates at which humans can move & \cite{weiss2018global}\\
LST day &  Daytime land surface temperature & \cite{modisLST} \\
Night Lights & Index that measures the presence of lights from towns, cities and other sites with persistent lighting & \cite{elvidge2017viirs} \\
CHIRPS & Climate Hazards Group Infrared Precipitation with Station Data; measure of precipitation & \cite{funk2014quasi} \\
Slope & Average slope of the pixel, determined from the elevation & \cite{farr2007shuttle}\\
TCB & Tasselled cap brightness; measure of land reflectance & \cite{modisTCB}\\
TCW & Tasselled cap wetness &  \cite{modisTCB}\\
TSI Pv & Temperature suitability index for \textit{Plasmodium vivax} & \cite{gething2011modelling}\\
TSI Pf & Temperature suitability index for \textit{Plasmodium facliparum} & \cite{gething2011modelling}\\
LST difference & Difference between day and night time land surface temperature &\cite{modisLST}
\end{tabular}
\end{table}

The real covariates used in this study are listed in Table \ref{cov_table}. Examples of these covariates are shown in Figure \ref{fig:ex_covs} and the full sets of covariates are shown in the supplementary material. These are a variety of environmental and socioeconomic variables that are among those commonly used for spatial mapping of malaria risk (see e.g. \cite{kang2018spatio, weiss2019mapping, battle2019mapping, arambepola2020nonparametric}). Many of the environmental factors affect mosquito breeding habitats (typically pools of stagnant water) or parasite development, while the socioeconomic variables may be correlated with access to healthcare or rural/urban settings. These covariates were chosen to provide a reasonable amount of variation in the spatial patterns and spatial scales at which they vary. In each country only the temperature suitability index of the dominant parasite \citep{gething2011modelling} was used (\textit{Vivax} in India and \textit{Falciparum} in Madagascar). In addition, the `Night Lights' variable, an index  that  measures  the  presence of  lights  from  towns,  cities  and other sites with persistent lighting \citep{elvidge2017viirs}, was used in India but there was very little variation in this variable in Madagascar outside of the capital Antananarivo, so in Madagascar it was replaced by the `LST difference' variable, the difference between day and night time land surface temperature \citep{modisLST} .
\begin{figure}
\centering
   \includegraphics[width=0.85\linewidth]{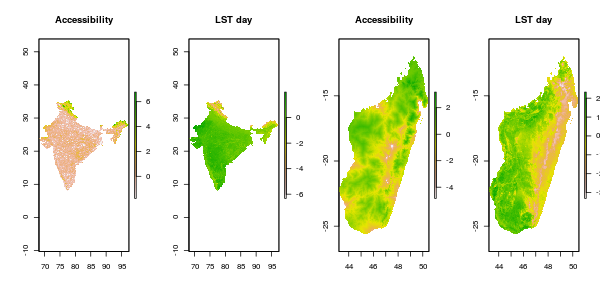}
   \caption{Two real covariates used in the study, accessibility to cities  and land surface daytime temperature in India (left) and Madagascar (right).}
   \label{fig:ex_covs} 
\end{figure}

The mock covariates were simulated M\'atern random fields with varying randomly sampled scales. These covariates generally varied on a shorter spatial scale than some of the real covariates and were designed to represent the generally unobserved socioeconomic and human factors that influence risk. Examples of these mock covariates are shown in Figure \ref{fig:ex_mock_covs} and shown in full in the supplementary material. It is worth noting that the sum of these simulated covariates is a Gaussian process but does not in general have a M\'atern covariance structure, and therefore the addition of multiple mock covariates is not equivalent to adding a single mock covariate with larger variance.
\begin{figure}
\centering
   \includegraphics[width=0.85\linewidth]{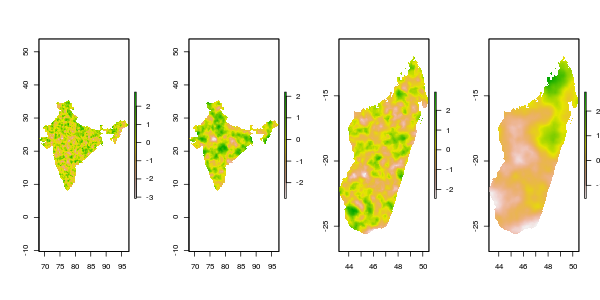}
   \caption{Examples of mock covariates for India (left) and Madagascar (right).}
   \label{fig:ex_mock_covs} 
\end{figure}

\subsection{Disaggregation regression model}
We assume that aggregated incidence data $y_1,...,y_N$ is available from $N$ spatial polygons (for example total case numbers in $N$ districts) and we wish to predict risk on a grid of pixels, each of which is contained in exactly one polygon. The number of cases in pixel $j$ of polygon $i$, $y_{ij}$, is modelled as a realisation of a Poisson random variable with mean
$$\hspace{4cm}\mu_{ij} = \lambda_{ij}\times p_{ij} \quad\quad i\in\{1,...,N\}, \,j\in\{1,...,M_i\}$$
where $\lambda_{ij}$ is the underlying risk in that pixel, $p_{ij}$ is the pixel population and $M_i$ is the number of pixels in polygon $i$. As is common in geostatistical models \citep{diggle1998model}, risk can be modelled as a smooth surface which, when transformed with suitable link function, is given by a the sum of a linear combination of covariates and a spatially correlated noise term
$$\log\lambda_{ij} = \beta_0 + \bm{\beta}^T\bm{X}_{ij} + \epsilon_{ij}.$$
Here the link function is $\log$, which is typical when modelling count data, and $\bm{X}_{ij}$ are the covariate values in this pixel. The spatial noise $\{\epsilon_{ij}\}$ is modelled as a realisation of a Gaussian process with a M\'atern covariance structure, parameterised by the range, $\rho$, and scale, $\sigma$. This spatial term can be thought of as representing factors which affect risk but have not (or cannot) be measured. The number of cases observed in polygon $i$, $y_i$, is assumed to be the sum of the (unobserved) number of cases in each pixel,
$$y_i = \sum_{j=1}^{M_i} y_{ij}.$$
Assuming that, conditional on the underlying risk surface, the Poisson processes in each pixel are independent, this sum also follows a Poisson distribution with mean equal to the sum of the means of each pixel process, i.e.
$$y_i \sim \mathrm{Pois}\left(\sum_{j=1}^{M_i}p_{ij}\lambda_{ij}\right)$$ 
which allows us to compute the likelihood of the model parameters $\beta_0\in\mathbb{R}$, $\bm{\beta}\in\mathbb{R}^K$ and the range and scale parameters of the M\'atern covariance function. The Bayesian model is completed by specifying priors on each model parameter. Gaussian priors were placed on $\beta_0$, with mean -4 and standard deviation 2, and $\bm{\beta}$, with mean 0 and standard deviation 1. Penalised complexity priors were used for the range and scale parameters of the spatial Gaussian process \citep{fuglstad2019constructing}, with a prior probability of 0.01 that $\rho > 1$ and 0.01 that $\sigma < 0.5$. The model was implemented in R using the disaggregation package \citep{nandi2019}. Parameters were estimated by maximising the posterior.

\FloatBarrier
\section{Evaluating fine-scale predictive performance}\label{exp1}
\subsection{Model fitting and prediction}
The performance of disaggregation regression was evaluated in the three misspecification scenarios with varying numbers of observations and sizes of polygons. For each administrative level and number of polygons, a contiguous set of polygons of this administrative level and number was randomly sampled (see Figure \ref{fig:shapes} for examples). The model was then fit using observations from these polygons (aggregated from the pixel case surface) and pixel-level incidence rates were predicted. For each country and scenario, this was repeated over all 20 simulated risk surfaces.

Pixel-level predicted and true incidence rates were compared only within polygons where response data was observed. The main metrics used were overall correlation and, to measure the ability of the model to predict risk patterns within polygons, correlation within each polygon. The performance of the model in terms of overall correlation was also compared to a baseline of simply assigning each pixel the observed polygon-level incidence rate.

\subsection{Results}
\begin{figure}
\centering
\includegraphics[width=0.9\linewidth]{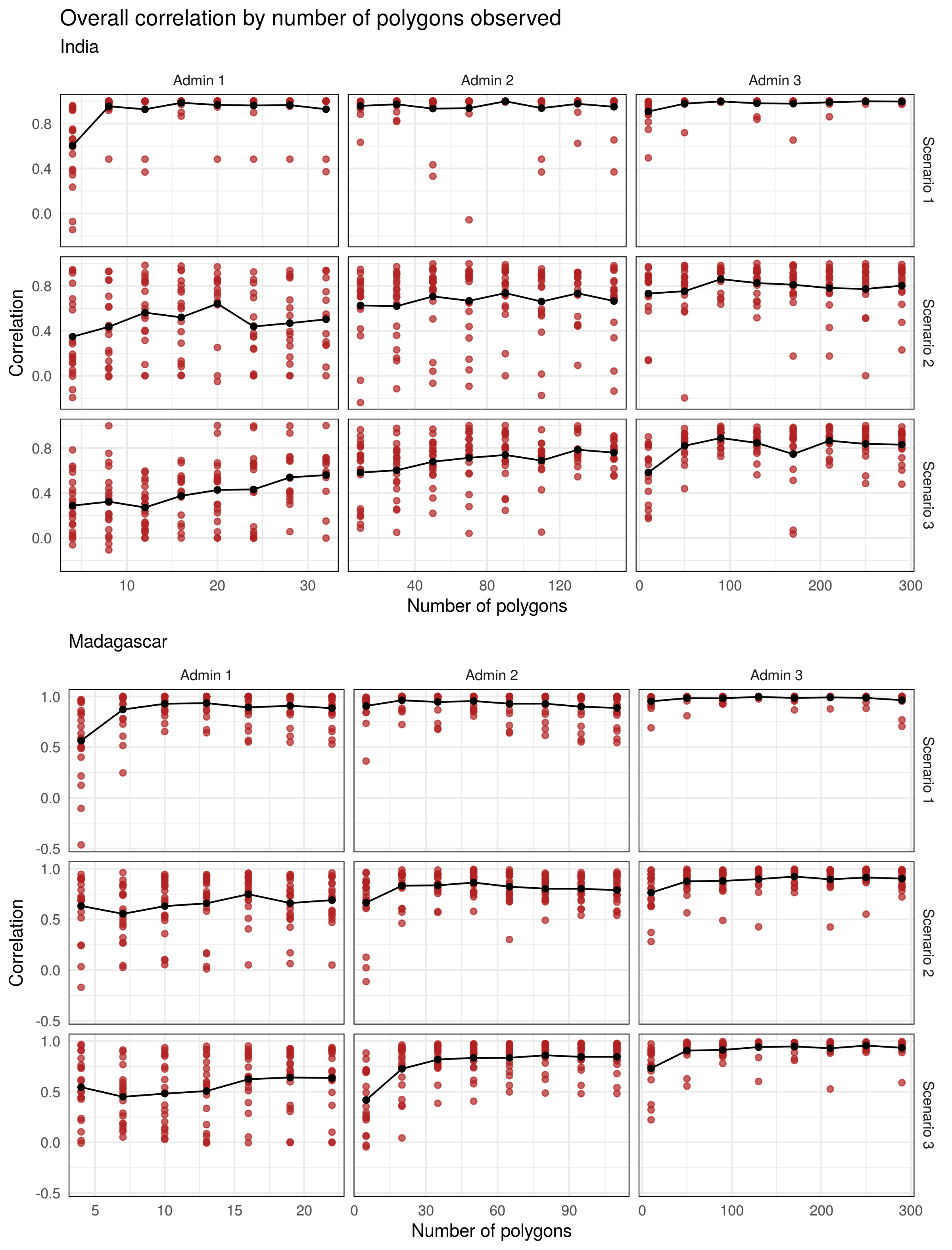}
  \caption{Overall correlation in India (top) and Madagascar (bottom) between observed and predicted pixel rates against number of polygons observed, stratified by scenario and administrative level. The mean over all repeats is given by the black line.}
  \label{fig:overall_cor}
\end{figure}

The overall correlation between observed and predicted pixel rates as the number of observations increases is shown in Figure \ref{fig:overall_cor}, stratified by scenario and administrative level. The overall trends were largely as expected, with more data points (number of polygons observed), smaller polygons and less model misspecification all generally resulting in improved pixel predictions. 


In scenario 1, where the model was well-specified, there was little difference in correlation as polygon size or number of polygons observed varied. This was particularly true in India where, with the exception of when the number of observations was very small, correlations were uniformly very high, whereas in Madagascar polygon size appeared to have more of an effect. In general it appears the relationships between covariates and risk were learned effectively, even for large polygons or relatively few data points, and therefore pixel-level predictions were consistently good.

In scenarios 2 and 3, where some of the covariates used to produce the data were withheld during the fitting process, there were clearer trends in performance as the number of observations and size of polygons varied. When using the largest polygons, administrative level 1, performance varied greatly between repeats and on average the correlation was fairly low. As polygon size decreased, overall correlation improved and predictions were more reliable. For administrative level 3 units, particularly in Madagascar, overall correlation was consistently high despite the model misspecification. In both countries and across administrative levels, there was an initial trend of improved performance as the number of observations increased which usually levelled off. Model performance in terms of overall correlation was typically better in scenario 2 than scenario 3, though the differences were often fairly small.
\begin{figure}
\centering
   \includegraphics[width=0.95\linewidth]{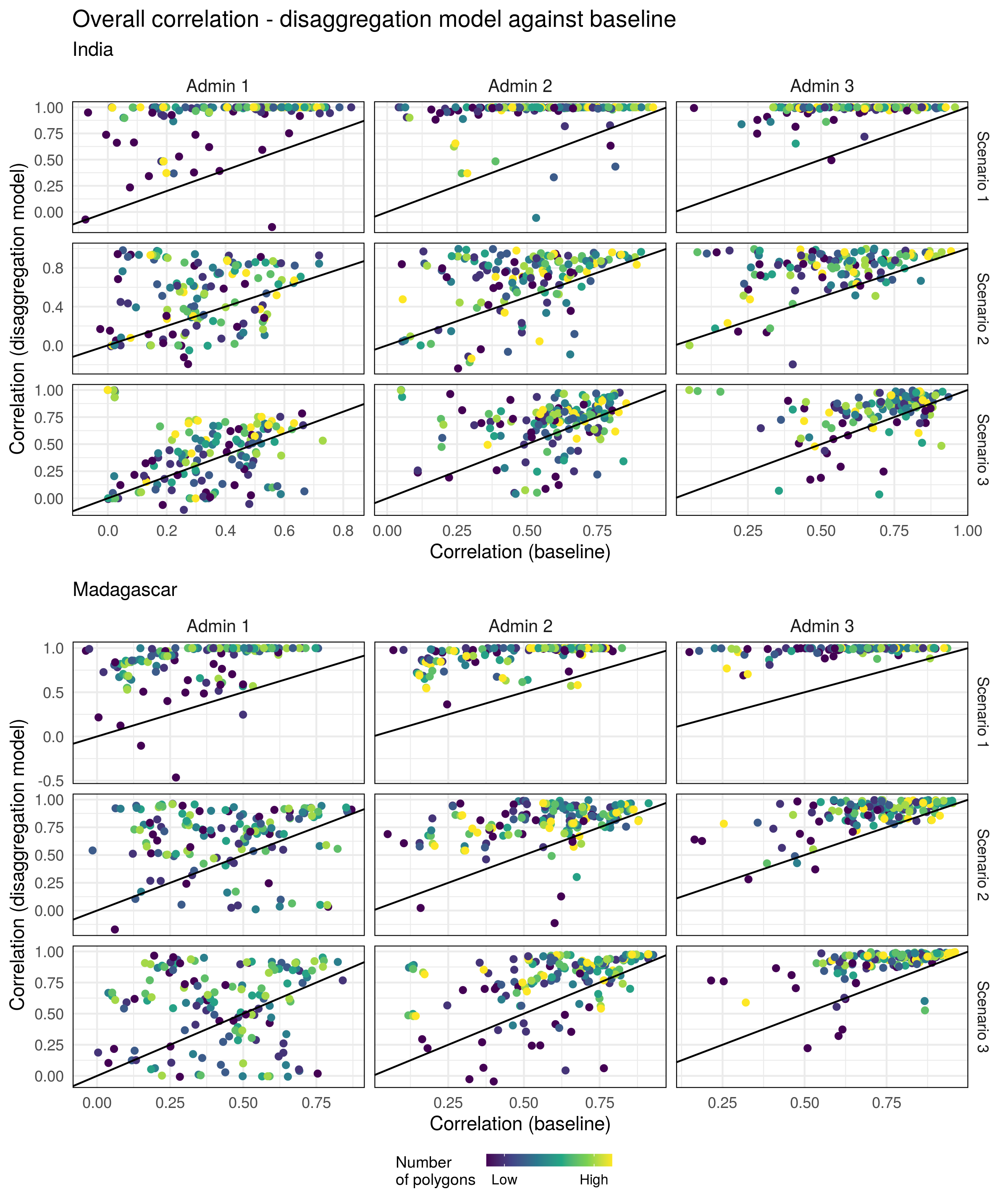}
  \caption{Comparison of correlation between predicted and observed rates when using the disaggregation model and when assigning the observed polygon rate to each pixel in that polygon, stratified by scenario and administrative unit size in India (top) and Madagascar (bottom), with the line $y=x$.}
  \label{fig:baseline}
\end{figure}

Figure \ref{fig:baseline} compares the disaggregation model with a baseline model which simply assigns the observed polygon rate to each pixel in that polygon. The disaggregation model generally greatly outperforms the baseline under no model misspecification. However, when there is model misspecification the results are more mixed. For administrative level 1, the disaggregation model sometimes performs worse than this simplistic baseline model (18\% and 35\% of the time in Madagascar and 35\% and 45\% of the time in India, in scenarios 2 and 3 respectively), even for large numbers of observations. When using administrative level 2 units, the disaggregation model typically outperforms the baseline by this metric, although for small numbers of polygons in scenario 3 it sometimes does worse. Using administrative level 3 units, the disaggregation model performs better in the large majority of cases although often the difference between methods is small.
\begin{figure}
\centering
  \includegraphics[width=0.9\linewidth]{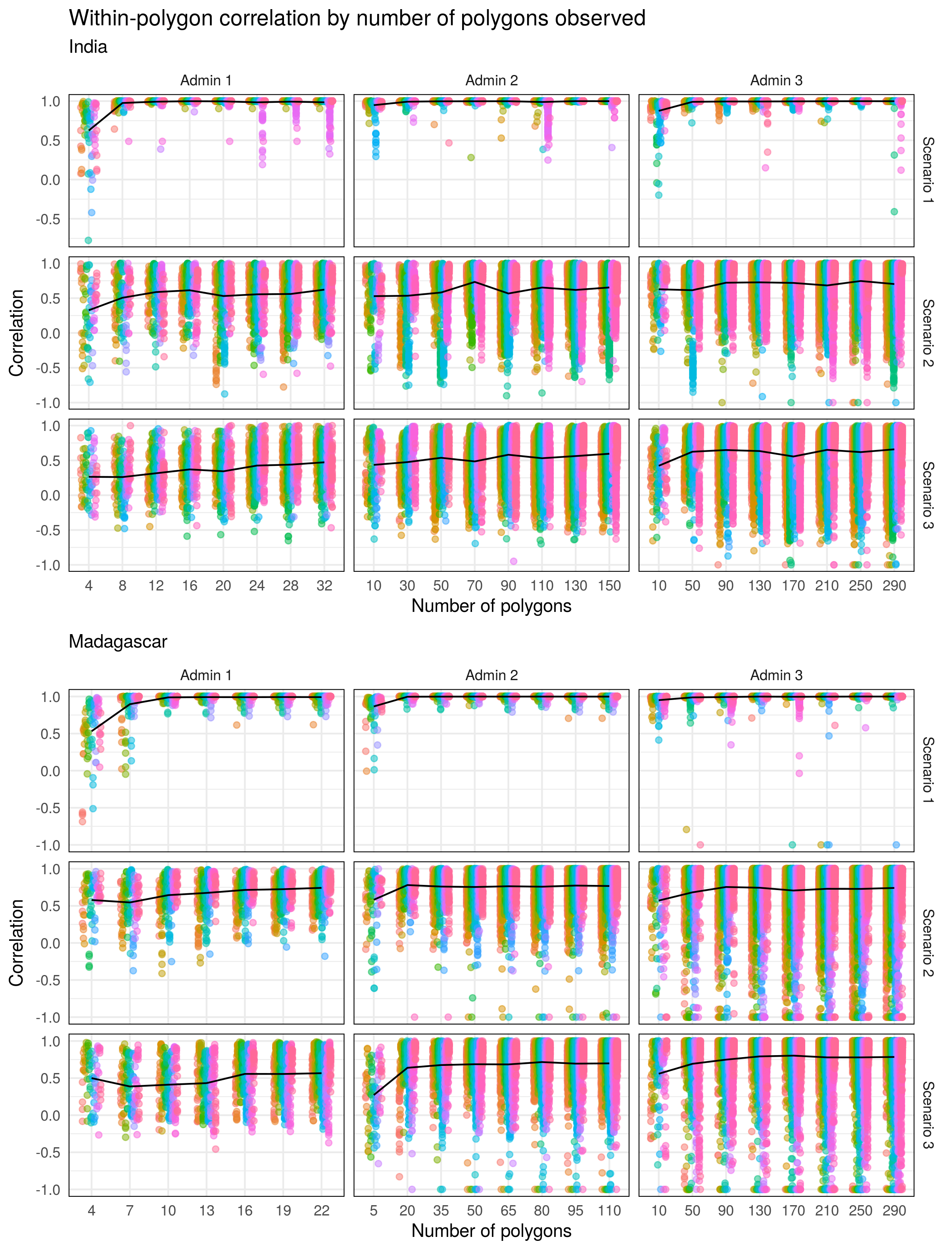}
  \caption{Correlation in India (top) and Madagascar (bottom) between observed and predicted pixel rates in each polygon against number of polygons observed, stratified by scenario and administrative level. The colour of the points represents the different repeats and have been spread horizontally for greater clarity. The mean over all repeats is given by the black line.}
  \label{fig:within_cor}
\end{figure}

\begin{figure}
\centering
   \includegraphics[width=0.9\linewidth]{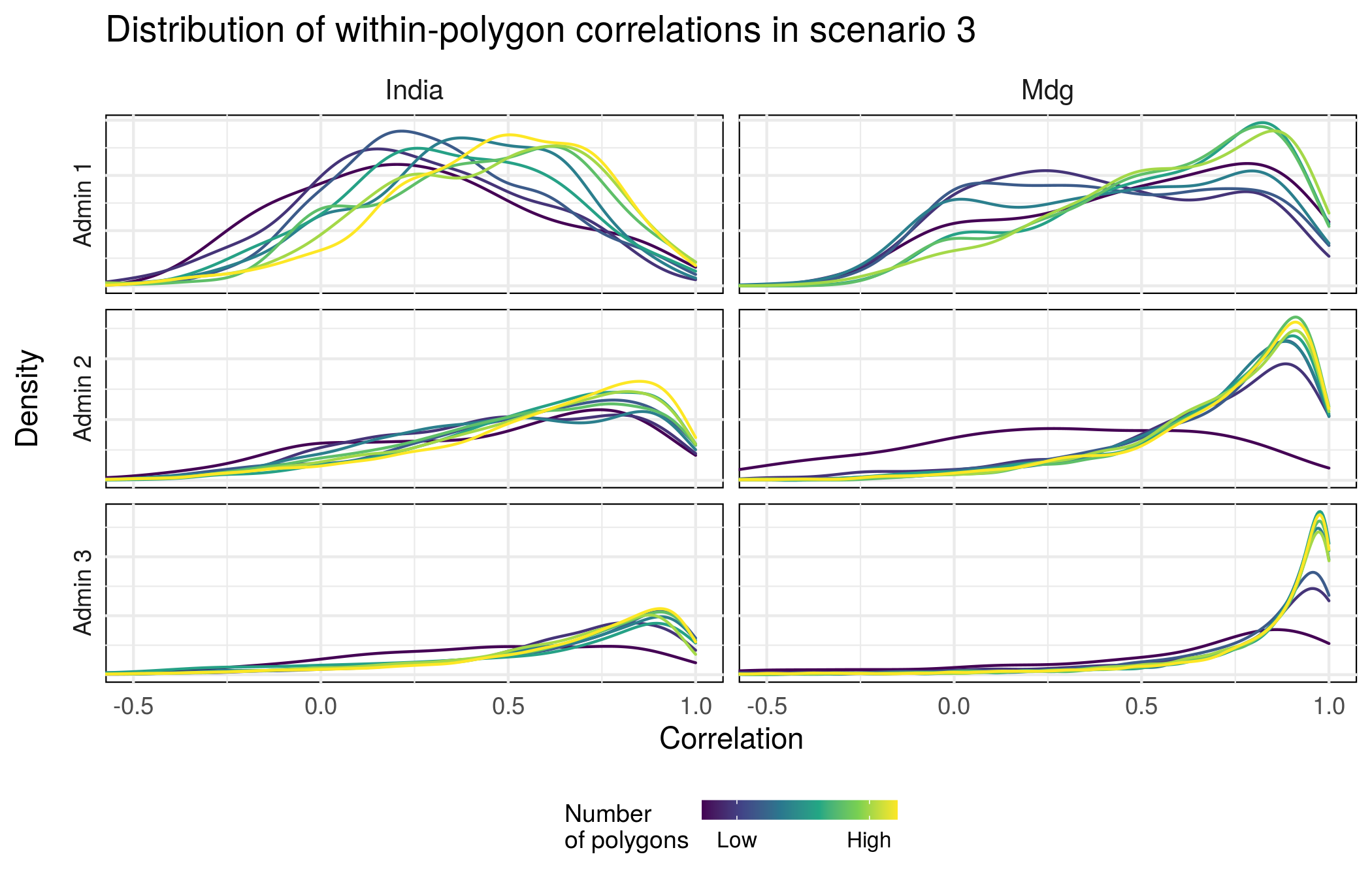}
  \caption{Distribution of correlations within each polygon in scenario 3 by administrative level and country.}
  \label{fig:cor_density}
\end{figure}

Figure \ref{fig:within_cor} shows the correlation between observed and predicted values within each polygon, which measures the ability of the model to predict heterogeneity in risk within each polygon. The mean of these within-polygon correlations shows a similar pattern to overall correlation, with uniformly good performance in scenario 1 and on average increasing correlation as number of observations increases and polygon size decreases in scenarios 2 and 3. However, when considering each polygon (rather than simply the average over all polygons and repeats), there is a large amount of heterogeneity in within-polygon performance when the model is misspecified (scenarios 2 and 3). These large variations are observed across all administrative levels and numbers of observations. Performance is also highly variable across different polygons in the same repeat, rather the model making consistently good or consistently poor predictions in some repeats. The distribution of these correlations in scenario 3 is shown in Figure \ref{fig:cor_density}. When using administrative level 1 units, there were a significant proportion of polygons with little to no correlation between predicted and true pixel rates, even for large numbers of observations, with a correlation of less than 0.3 in 27\% of polygons in Madagascar and 38\% in India. In India when using administrative level 2 units, correlations were generally high but there was still considerable variation. Correlations were mostly high for administrative level 2 in Madagascar and level 3 in India and consistently high for administrative level 3 in Madagascar.

\begin{figure}
\centering
   \includegraphics[width=0.8\linewidth]{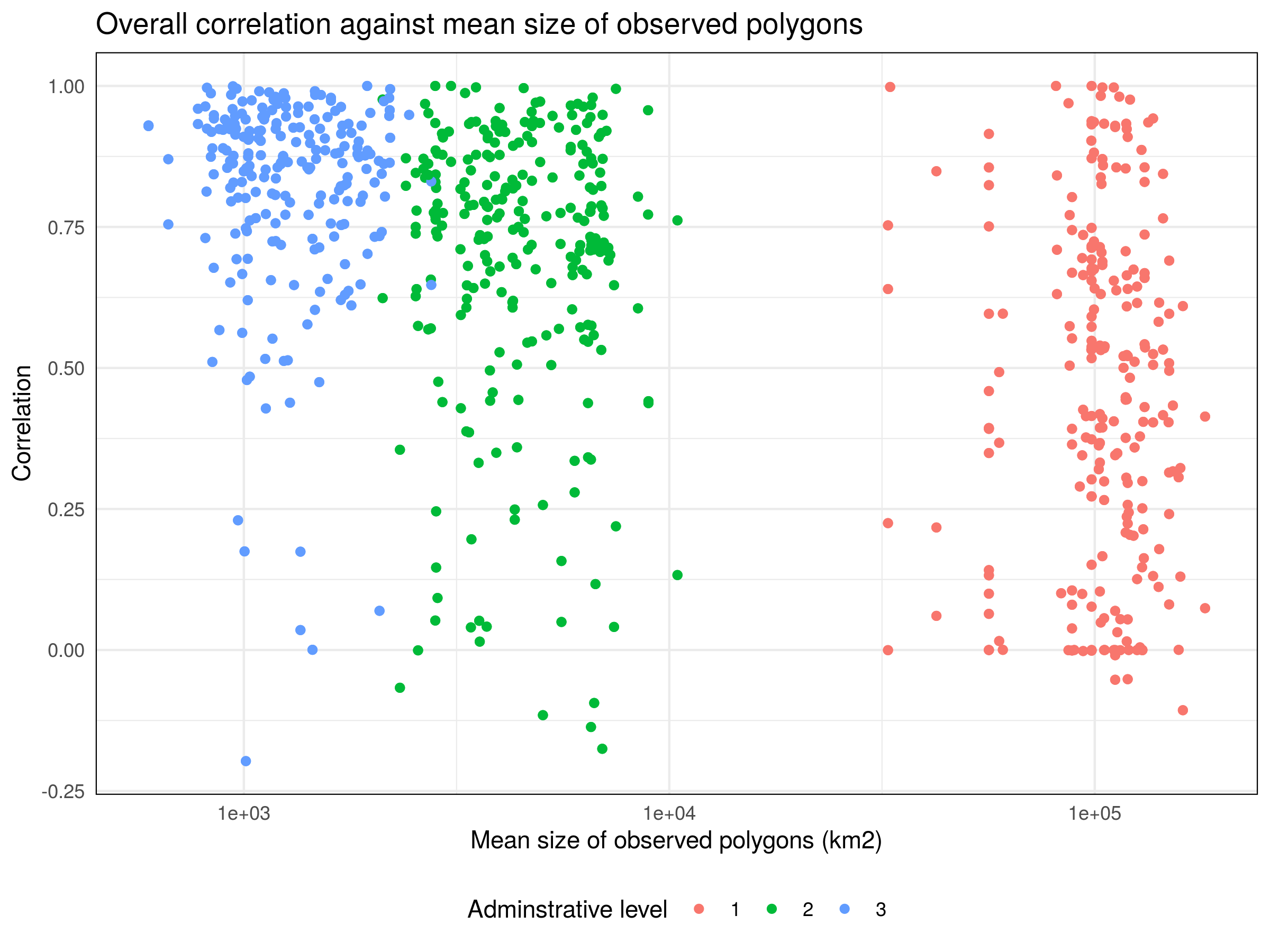}
  \caption{Overall correlation between observed and predicted rates against the mean size of observed polygons (on a log scale) in both India and Madagascar in scenarios 2 and 3.} 
    \label{fig:size_cor}
\end{figure}

The relationship between average size of observed polygons and overall correlation under model misspecification (scenarios 2 and 3) is shown in Figure \ref{fig:size_cor}. The average size varied between repeats even when the same administrative level was used because for each repeat a random set of contiguous polygons was chosen. The negative association between polygon size and predictive performance previously observed when comparing administrative levels also appears to hold for different sets of polygon within the same administrative level, with a correlation of -0.44 between overall correlation and mean size of polygon. This negative relationship is strongest for administrative levels 2 and 3 while it is less clear in administrative level 1.

\section{Evaluating polygon-level cross validation as a metric of fine-scale predictive performance}\label{exp2}
\subsection{Model fitting and prediction}
The relationship between fine-scale predictive performance and polygon-level cross validation was investigated under scenario 3 (the highest level of misspecification), which we considered the most realistic scenario. We did not consider settings where, according to our results in the previous section, disaggregation regression was unlikely to be successfully and reliably applied. Therefore only administrative levels 2 and 3 were considered, with a minimum of 50 observations.

For each risk surface and administrative level, a contiguous set of polygons of a random size was sampled. The disaggregation model was applied over this area and the correlation between predicted and true pixel-level rates was calculated. We then performed five-fold cross validation at the polygon level over this area: The set of polygons was randomly split into five subsets of approximately equal size. For each of these subsets, response data from this subset was withheld and predictions for the rate in these held-out polygons were made by applying the disaggregation model with the remaining response data and aggregating pixel-level predictions in the held-out subset. The process was repeated for all subsets and the correlation between polygon predicted and true rates was calculated.

This was repeated five times for each risk surface and the pixel-level and cross-validated polygon-level correlations were compared.

\subsection{Results}
\begin{figure}
\centering
  \includegraphics[width=0.8\linewidth]{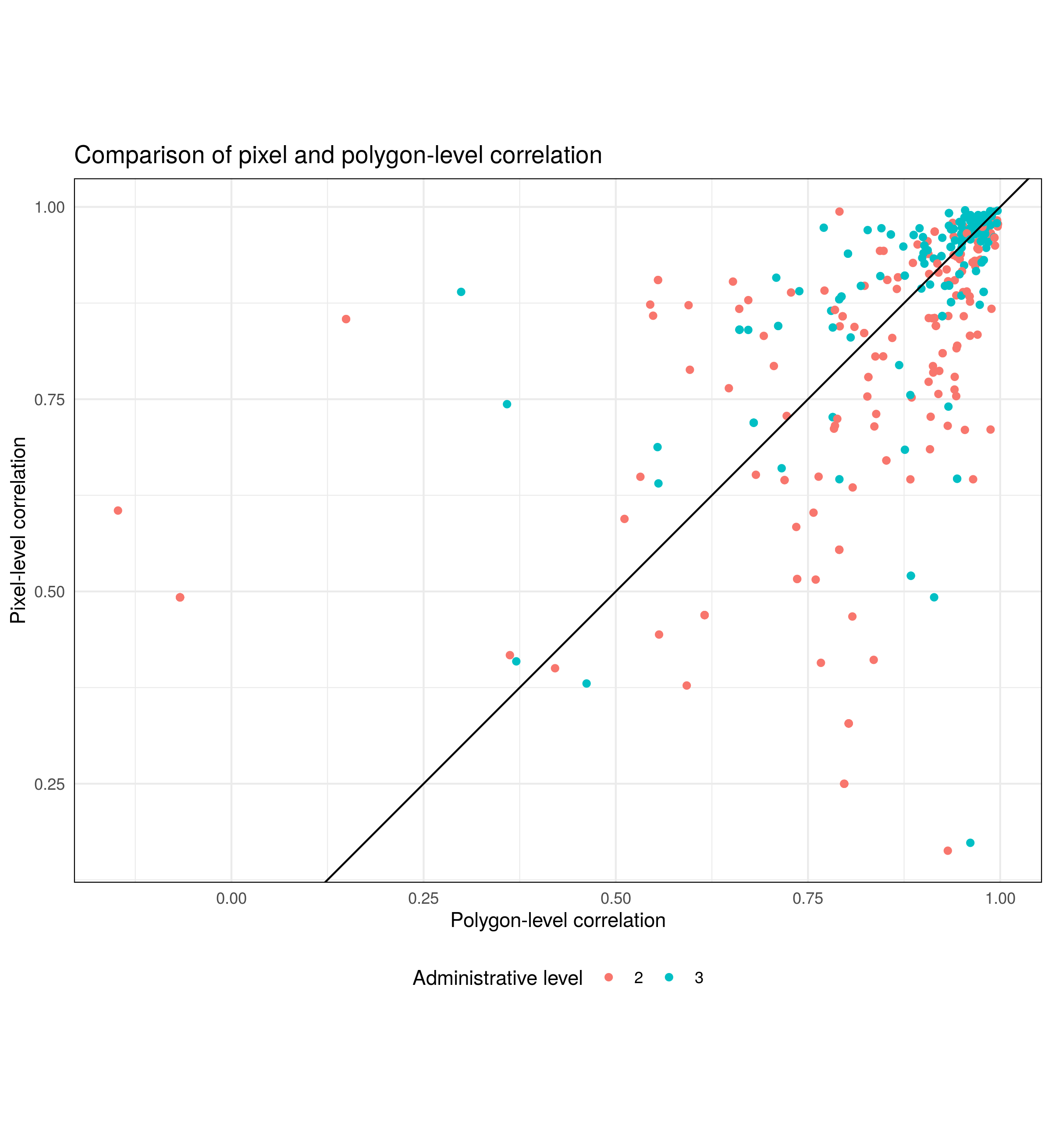}
  \caption{Comparison of correlation between predicted and observed polygon rates with five-fold cross validation and predicted and true pixel rates in full model fit.}
  \label{fig:cv_results}
\end{figure}

The comparison between correlation on the pixel-level and aggregated cross-validated correlation is shown in Figure \ref{fig:cv_results}. Overall, there was a moderately strong relationship, with a correlation of 0.52 between these two metrics. The polygon correlation was greater than pixel correlation more often when administrative level 2 units were used (69\% of the time) compared to administrative level 3 units (44\% of the time). When considered separately, the strength of the relationships between the metrics was slightly higher for administrative level 3 units than administrative level 2 units, with correlations of 0.53 and 0.47 respectively. 

Despite the positive association between pixel and polygon-level metrics, however, there still were a number of repeats (usually when using administrative level 2 units) where polygon-level correlation was high but pixel-level correlation was much lower. For example,  of repeats in which pixel-level correlation was less than 0.6 (representing roughly the worst 10\% of pixel-level correlations), in more than half (53\%) polygon-level correlation was above 0.75.

\FloatBarrier

\section{Discussion}
While we included model misspecification in this study in the form of missing covariates, there are many other factors that we did not consider when generating mock incidence data. These include non-linear covariate interactions, incomplete reporting, treatment seeking behaviour, human mobility and alternative case-generating processes. Although it may be possible account for some of these factors, it therefore seems fairly likely that, even compared to our most misspecified scenario, the performance of disaggregation regression could be worse when applied to real data than in this study.

The results in Section \ref{exp1} demonstrate that disaggregation regression works well under no model misspecification, with relationships between the covariates and risk learned effectively (avoiding the ecological fallacy), even with relatively few polygons and relatively large polygons. However, this is a fairly unrealistic scenario. In addition to the factors not included in the generative model mentioned above, it is unlikely that all variables that significantly influence disease risk can be measured, let alone that these values will be available on a fine-scale grid across the region of interest. Therefore the results of this scenario serve as a proof of concept but do little to inform real world risk mapping problems. 

Scenarios 2 and 3 represent more realistic situations and the results here suggest that there may be little benefit in applying disaggregation regression when the observed data is aggregated over large areas. For administrative level 1 units, model performance was highly variable and the spatial patterns predicted by the disaggregation model were often less accurate than using the observed aggregate rates. On average the model was able to capture some sub-polygon heterogeneity accurately but again the large variation in performance between polygons and repeats limits the practical use of these predictions. While performance generally improved as the number of observations increased, these issues were still present with large numbers of observations and the maximum number of observations will necessarily be limited by the size of the study area. Administrative level 2 units in India may also be too large to reliably make fine-scale predictions from real data.

However, for smaller polygons (administrative level 2 in Madagascar and level 3 in both countries) the results of this study show that fine-scale patterns can be accurately inferred from aggregated data. In these cases, the fine-scale predictions made using the disaggregation model captured both overall and within-polygon heterogeneity well and therefore were a significant improvement over the observed aggregated data. Performance was also much more reliable, both between repeats and (when considering within-polygon correlation) between polygons. Performance improved as the number of observations increased but generally high correlations were reached with a moderate number of observations. Largely as expected, performance was better when aggregated areas were smaller, a trend which was observed both between and within administrative levels. Fine-scale predictions were particularly good for administrative level 3 in Madagascar.

The results from Section \ref{exp2} suggest that cross-validation on a polygon-level has some use as a proxy for fine-scale predictive performance. There was a correlation of around 0.5 between polygon and pixel-level metrics and therefore a high polygon correlation should increase confidence in fine-scale predictions. However, the relationship between polygon and pixel-level correlation may still not be strong enough to consistently identify poor pixel-level predictions from polygon-level correlation. This was the case in our simulations, as polygon-level correlation was often high even when pixel-level correlation was low. It therefore remains important to validate disaggregation regression using fine-scale observations wherever possible. Where no fine-scale data exists, alternative metrics of similar phenomena may help evaluate fine-scale patterns qualitatively. 


\section{Conclusion}
The results of this study suggest that disaggregation regression can be applied successfully in applications where response data is aggregated over small areas. The different situations considered in this study (in terms of polygon size, number of observations and level of misspecification) should allow for more informed decisions to be made about whether disaggregation regression is an appropriate method for a specific problem.

However, caution should be taken when response data is aggregated over large areas. While disaggregation was on average moderately effective when using large polygons, predictions were generally unreliable, with large variations in performance both between repeats and between regions in the same repeat. Our results suggest that administrative level 1 and 2 units in India and level 1 units in Madagascar may be too large to perform effective disaggregation. In these cases it may be more useful to model on the aggregate level and look for alternative sources of data (such as data from sub-regions) to inform sub-polygon heterogeneity.

In our simulations, cross-validation correlation at the aggregate level was correlated with fine-scale accuracy and therefore in real world applications correlation on the aggregate level may be of some use as a validation metric where only aggregated data is available. However, a high correlation on the aggregate level is not a guarantee of accurate fine-scale predictions and wherever possible Fine-scale predictions should be validated using data on the same spatial scale. 


\section*{Acknowledgments}
The first author was supported in this work through an Engineering and Physical Sciences Research Council (EPSRC) (https://epsrc.ukri.org/) Systems Biology studentship
award \\(EP/G03706X/1). Work by the Malaria Atlas Project on methods development
for Malaria Eradication Metrics including this work is supported by a grant from the
Bill and Melinda Gates Foundation (OPP1197730).

\bibliographystyle{biorefs}  
\bibliography{refs}

\begin{thebibliography}{99}

\bibitem[Arambepola \emph{and others}(2020)Arambepola, Gething and
  Cameron]{arambepola2020nonparametric}
\textsc{Arambepola, Rohan, Gething, Peter and Cameron, Ewan}. (2020).
\newblock Nonparametric causal feature selection for spatiotemporal risk
  mapping of malaria incidence in {Madagascar}.
\newblock {\em arXiv preprint arXiv:2001.07745\/}.

\bibitem[Battle \emph{and others}(2019)Battle, Lucas, Nguyen, Howes, Nandi,
  Twohig, Pfeffer, Cameron, Rao, Casey, Bertozzi-Villa, Collins, Dalrymple,
  Gray, Harris, Howes, Keddie, May, Rumisha, Thorn, Barber, Fullman, Huynh,
  Kulikoff, Kutz, Lope, Mokdad, Naghavi, Nguyen, Shackelford, Vos, Wang, Smith,
  Lim, Murray, Bhatt, Weiss, Hay and Gething]{battle2019mapping}
\textsc{Battle, Katherine~E, Lucas, Tim~CD, Nguyen, Michele, Howes, Rosalind~E,
  Nandi, Anita~K, Twohig, Katherine~A, Pfeffer, Daniel~A, Cameron, Ewan, Rao,
  Puja~C, Casey, Daniel, Bertozzi-Villa, Amelia, Collins, Emma~L, Dalrymple,
  Ursula, Gray, Naomi, Harris, Joseph~R, Howes, Rosalind~E, Keddie, Suzanne~H,
  May, Daniel, Rumisha, Susan, Thorn, Michael~P, Barber, Ryan, Fullman, Nancy,
  Huynh, Chantal~K, Kulikoff, Xie, Kutz, Michael~J, Lope, Alan~D, Mokdad,
  Ali~H, Naghavi, Mohsen, Nguyen, Grant, Shackelford, Katya~Anne, Vos, Theo,
  Wang, Haidong, Smith, David~L, Lim, Stephen~S, Murray, Christopher J~L,
  Bhatt, Samir, Weiss, Daniel~J, Hay, Simon~I} \emph{and others}. (2019).
\newblock Mapping the global endemicity and clinical burden of {Plasmodium}
  vivax, 2000--17: a spatial and temporal modelling study.
\newblock {\em The Lancet\/}~\textbf{394}(10195), 332--343.

\bibitem[Diggle \emph{and others}(2003)Diggle, Ribeiro and
  Christensen]{diggle2003introduction}
\textsc{Diggle, Peter~J, Ribeiro, Paulo~J and Christensen, Ole~F}. (2003).
\newblock An introduction to model-based geostatistics.
\newblock In:  {\em Spatial statistics and computational methods\/}. Springer,
  pp.\  43--86.

\bibitem[Diggle \emph{and others}(1998)Diggle, Tawn and
  Moyeed]{diggle1998model}
\textsc{Diggle, Peter~J, Tawn, Jonathan~A and Moyeed, RA}. (1998).
\newblock Model-based geostatistics.
\newblock {\em Journal of the Royal Statistical Society: Series C (Applied
  Statistics)\/}~\textbf{47}(3), 299--350.

\bibitem[Drake \emph{and others}(2017)Drake, Lubell, Kyaw, Devine, Kyaw, Day,
  Smithuis and White]{drake2017geographic}
\textsc{Drake, Tom~L, Lubell, Yoel, Kyaw, Shwe~Sin, Devine, Angela, Kyaw,
  Myat~Phone, Day, Nicholas~PJ, Smithuis, Frank~M and White, Lisa~J}. (2017).
\newblock Geographic resource allocation based on cost effectiveness: an
  application to malaria policy.
\newblock {\em Applied health economics and health policy\/}~\textbf{15}(3),
  299--306.

\bibitem[Elliot \emph{and others}(2000)Elliot, Wakefield, Best and
  Briggs]{elliot2000spatial}
\textsc{Elliot, Paul, Wakefield, Jon~C, Best, Nicola~G and Briggs, David~John}.
  (2000).
\newblock {\em Spatial epidemiology: methods and applications.\/}. Oxford
  University Press.

\bibitem[Elvidge \emph{and others}(2017)Elvidge, Baugh, Zhizhin, Hsu and
  Ghosh]{elvidge2017viirs}
\textsc{Elvidge, Christopher~D, Baugh, Kimberly, Zhizhin, Mikhail, Hsu,
  Feng~Chi and Ghosh, Tilottama}. (2017).
\newblock {VIIRS} night-time lights.
\newblock {\em International Journal of Remote Sensing\/}~\textbf{38}(21),
  5860--5879.

\bibitem[Farr \emph{and others}(2007)Farr, Rosen, Caro, Crippen, Duren,
  Hensley, Kobrick, Paller, Rodriguez, Roth, Seal, Shaffer, Shimada, Umland,
  Werner, Oskin, Burbank and Alsdorf]{farr2007shuttle}
\textsc{Farr, Tom~G, Rosen, Paul~A, Caro, Edward, Crippen, Robert, Duren,
  Riley, Hensley, Scott, Kobrick, Michael, Paller, Mimi, Rodriguez, Ernesto,
  Roth, Ladislav, Seal, David, Shaffer, Scott, Shimada, Joanne, Umland,
  Jeffrey, Werner, Marian, Oskin, Michael, Burbank, Douglas} \emph{and others}.
  (2007).
\newblock The shuttle radar topography mission.
\newblock {\em Reviews of geophysics\/}~\textbf{45}(2).

\bibitem[Fotheringham and Wong(1991)Fotheringham and
  Wong]{fotheringham1991modifiable}
\textsc{Fotheringham, A~Stewart and Wong, David~WS}. (1991).
\newblock The modifiable areal unit problem in multivariate statistical
  analysis.
\newblock {\em Environment and planning A\/}~\textbf{23}(7), 1025--1044.

\bibitem[Fuglstad \emph{and others}(2019)Fuglstad, Simpson, Lindgren and
  Rue]{fuglstad2019constructing}
\textsc{Fuglstad, Geir-Arne, Simpson, Daniel, Lindgren, Finn and Rue,
  H{\aa}vard}. (2019).
\newblock Constructing priors that penalize the complexity of {Gaussian} random
  fields.
\newblock {\em Journal of the American Statistical
  Association\/}~\textbf{114}(525), 445--452.

\bibitem[Funk \emph{and others}(2014)Funk, Peterson, Landsfeld, Pedreros,
  Verdin, Rowland, Romero, Husak, Michaelsen and Verdin]{funk2014quasi}
\textsc{Funk, Chris~C, Peterson, Pete~J, Landsfeld, Martin~F, Pedreros,
  Diego~H, Verdin, James~P, Rowland, James~D, Romero, Bo~E, Husak, Gregory~J,
  Michaelsen, Joel~C and Verdin, Andrew~P}. (2014).
\newblock A quasi-global precipitation time series for drought monitoring.
\newblock {\em US Geological Survey Data Series\/}~\textbf{832}(4), 1--12.

\bibitem[Gething \emph{and others}(2011)Gething, Van~Boeckel, Smith, Guerra,
  Patil, Snow and Hay]{gething2011modelling}
\textsc{Gething, Peter~W, Van~Boeckel, Thomas~P, Smith, David~L, Guerra,
  Carlos~A, Patil, Anand~P, Snow, Robert~W and Hay, Simon~I}. (2011).
\newblock Modelling the global constraints of temperature on transmission of
  {Plasmodium falciparum and P. vivax}.
\newblock {\em Parasites \& vectors\/}~\textbf{4}(1), 92.

\bibitem[Kang \emph{and others}(2018)Kang, Battle, Gibson, Ratsimbasoa,
  Randrianarivelojosia, Ramboarina, Zimmerman, Weiss, Cameron, Gething,  and
  Howes]{kang2018spatio}
\textsc{Kang, Su~Yun, Battle, Katherine~E, Gibson, Harry~S, Ratsimbasoa,
  Ars{\`e}ne, Randrianarivelojosia, Milijaona, Ramboarina, St{\'e}phanie,
  Zimmerman, Peter~A, Weiss, Daniel~J, Cameron, Ewan, Gething, Peter~W, }
  \emph{and others}. (2018).
\newblock Spatio-temporal mapping of {Madagascar’s Malaria Indicator Survey}
  results to assess {Plasmodium} falciparum endemicity trends between 2011 and
  2016.
\newblock {\em BMC medicine\/}~\textbf{16}(1), 71.

\bibitem[Keil \emph{and others}(2013)Keil, Belmaker, Wilson, Unitt and
  Jetz]{keil2013downscaling}
\textsc{Keil, Petr, Belmaker, Jonathan, Wilson, Adam~M, Unitt, Philip and Jetz,
  Walter}. (2013).
\newblock Downscaling of species distribution models: a hierarchical approach.
\newblock {\em Methods in Ecology and Evolution\/}~\textbf{4}(1), 82--94.

\bibitem[Law \emph{and others}(2018)Law, Sejdinovic, Cameron, Lucas, Flaxman,
  Battle and Fukumizu]{law2018variational}
\textsc{Law, Ho~Chung, Sejdinovic, Dino, Cameron, Ewan, Lucas, Tim, Flaxman,
  Seth, Battle, Katherine and Fukumizu, Kenji}. (2018).
\newblock Variational learning on aggregate outputs with {Gaussian} processes.
\newblock In:  {\em Advances in Neural Information Processing Systems\/}. pp.\
  6081--6091.

\bibitem[Lawson \emph{and others}(1999)Lawson, Biggeri, B{\"o}hning, Lesaffre,
  Viel and Bertollini]{lawson1999disease}
\textsc{Lawson, Andrew, Biggeri, Annibale, B{\"o}hning, Dankmar, Lesaffre,
  Emmanuel, Viel, Jean-Francois and Bertollini, Roberto}. (1999).
\newblock {\em Disease mapping and risk assessment for public health\/}. Wiley
  New York.

\bibitem[Li \emph{and others}(2012)Li, Brown, Gesink and Rue]{li2012log}
\textsc{Li, Ye, Brown, Patrick, Gesink, Dionne~C and Rue, H{\aa}vard}. (2012).
\newblock Log {Gaussian Cox} processes and spatially aggregated disease
  incidence data.
\newblock {\em Statistical methods in medical research\/}~\textbf{21}(5),
  479--507.

\bibitem[Lucas \emph{and others}(2020)Lucas, Nandi, Chestnutt, Twohig, Keddie,
  Collins, Howes, Nguyen, Rumisha, Python, Arambepola, Bertozzi-Villa, Hancock,
  Amratia, Battle, Cameron, Gething and Weiss]{lucas2020mapping}
\textsc{Lucas, Tim~CD, Nandi, Anita~K, Chestnutt, Elisabeth~G, Twohig,
  Katherine~A, Keddie, Suzanne~H, Collins, Emma~L, Howes, Rosalin~E, Nguyen,
  Michele, Rumisha, Susan, Python, Andre, Arambepola, Rohan, Bertozzi-Villa,
  Amelia, Hancock, Penelope, Amratia, Punam, Battle, Katherine~E, Cameron,
  Ewan, Gething, Peter~W} \emph{and others}. (2020).
\newblock Mapping malaria by sharing spatial information between incidence and
  prevalence datasets.
\newblock {\em medRxiv\/}.

\bibitem[Nandi \emph{and others}(2019)Nandi, Lucas, Arambepola and
  Python]{nandi2019}
\textsc{Nandi, Anita, Lucas, Tim, Arambepola, Rohan and Python, Andre}. (2019).
\newblock {\em disaggregation: Disaggregation Modelling\/}.
\newblock R package version 0.1.2.

\bibitem[{NASA Earth Data}(2017{\em a}){NASA Earth Data}]{modisTCB}
\textsc{{NASA Earth Data}}. (2017{\em a}).
\newblock Land processes distributed active archive center..
\newblock
  \url{https://lpdaac.usgs.gov/dataset_discovery/modis/modis_products_table/mcd43b5}.
\newblock [Accessed Sept 2017].

\bibitem[{NASA Earth Data}(2017{\em b}){NASA Earth Data}]{modisEVI}
\textsc{{NASA Earth Data}}. (2017{\em b}).
\newblock {MODIS (MOD 13) - Gridded} vegetation indices {(NDVI and EVI)}.
\newblock
  \url{http://modis.gsfc.nasa.gov/data/dataprod/dataproducts.php?MOD_NUMBER=13}.
\newblock [Accessed Sept 2017].

\bibitem[{NASA Earth Observations}(2017){NASA Earth Observations}]{modisLST}
\textsc{{NASA Earth Observations}}. (2017).
\newblock Average land surface temperature.
\newblock
  \url{http://neo.sci.gsfc.nasa.gov/view.php?datasetId=MOD_LSTD_CLIM_M}.
\newblock [Accessed Sept 2017].

\bibitem[Pfeffer \emph{and others}(2018)Pfeffer, Lucas, May, Harris, Rozier,
  Twohig, Dalrymple, Guerra, Moyes, Thorn, Nguyen, Bhatt, Cameron, Weiss~J,
  Howes~E, Katherine~E, Gibson and Gething]{pfeffer2018malariaatlas}
\textsc{Pfeffer, Daniel~A, Lucas, Timothy~CD, May, Daniel, Harris, Joseph,
  Rozier, Jennifer, Twohig, Katherine~A, Dalrymple, Ursula, Guerra, Carlos~A,
  Moyes, Catherine~L, Thorn, Mike, Nguyen, Michele, Bhatt, Samir, Cameron,
  Ewan, Weiss~J, Daniel, Howes~E, Rosalind, Katherine~E, Battle, Gibson,
  Harry~S} \emph{and others}. (2018).
\newblock {malariaAtlas: an R interface to global malariometric data hosted by
  the Malaria Atlas Project}.
\newblock {\em Malaria journal\/}~\textbf{17}(1), 1--10.

\bibitem[Sturrock \emph{and others}(2014)Sturrock, Cohen, Keil, Tatem,
  Le~Menach, Ntshalintshali, Hsiang and Gosling]{sturrock2014fine}
\textsc{Sturrock, Hugh~JW, Cohen, Justin~M, Keil, Petr, Tatem, Andrew~J,
  Le~Menach, Arnaud, Ntshalintshali, Nyasatu~E, Hsiang, Michelle~S and Gosling,
  Roland~D}. (2014).
\newblock Fine-scale malaria risk mapping from routine aggregated case data.
\newblock {\em Malaria journal\/}~\textbf{13}(1), 421.

\bibitem[Taylor \emph{and others}(2018)Taylor, Andrade-Pacheco and
  Sturrock]{taylor2018continuous}
\textsc{Taylor, Benjamin~M, Andrade-Pacheco, Ricardo and Sturrock, Hugh~JW}.
  (2018).
\newblock Continuous inference for aggregated point process data.
\newblock {\em Journal of the Royal Statistical Society: Series A (Statistics
  in Society)\/}~\textbf{181}(4), 1125--1150.

\bibitem[Trabucco and Zomer(2009)Trabucco and Zomer]{trabucco2009global}
\textsc{Trabucco, Antonio and Zomer, Robert~J}. (2009).
\newblock Global aridity index (global-aridity) and global potential
  evapo-transpiration {(global-PET)} geospatial database.
\newblock {\em CGIAR Consortium for Spatial Information\/}.

\bibitem[Wakefield and Shaddick(2006)Wakefield and
  Shaddick]{wakefield2006health}
\textsc{Wakefield, Jon and Shaddick, Gavin}. (2006).
\newblock Health-exposure modeling and the ecological fallacy.
\newblock {\em Biostatistics\/}~\textbf{7}(3), 438--455.

\bibitem[Weiss \emph{and others}(2019)Weiss, Lucas, Nguyen, Nandi, Bisanzio,
  Battle, Cameron, Twohig, Pfeffer, Rozier, Gibson, Rao, Casey, Bertozzi-Villa,
  Collins, Dalrymple, Gray, Harris, Howes, Keddie, May, Rumisha, Thorn, Barber,
  Fullman, Huynh, Kulikoff, Kutz, Lope, Mokdad, Naghavi, Nguyen, Shackelford,
  Vos, Wang, Smith, Lim, Murray, Bhatt, Hay and Gething]{weiss2019mapping}
\textsc{Weiss, Daniel~J, Lucas, Tim~CD, Nguyen, Michele, Nandi, Anita~K,
  Bisanzio, Donal, Battle, Katherine~E, Cameron, Ewan, Twohig, Katherine~A,
  Pfeffer, Daniel~A, Rozier, Jennifer~A, Gibson, Harry~S, Rao, Puja~C, Casey,
  Daniel, Bertozzi-Villa, Amelia, Collins, Emma~L, Dalrymple, Ursula, Gray,
  Naomi, Harris, Joseph~R, Howes, Rosalind~E, Keddie, Suzanne~H, May, Daniel,
  Rumisha, Susan, Thorn, Michael~P, Barber, Ryan, Fullman, Nancy, Huynh,
  Chantal~K, Kulikoff, Xie, Kutz, Michael~J, Lope, Alan~D, Mokdad, Ali~H,
  Naghavi, Mohsen, Nguyen, Grant, Shackelford, Katya~Anne, Vos, Theo, Wang,
  Haidong, Smith, David~L, Lim, Stephen~S, Murray, Christopher J~L, Bhatt,
  Samir, Hay, Simon~I} \emph{and others}. (2019).
\newblock Mapping the global prevalence, incidence, and mortality of
  {Plasmodium} falciparum, 2000--17: a spatial and temporal modelling study.
\newblock {\em The Lancet\/}.

\bibitem[Weiss \emph{and others}(2018)Weiss, Nelson, Gibson, Temperley,
  Peedell, Lieber, Hancher, Poyart, Belchior, Fullman, Mappin, Dalrymple,
  Rozier, Lucas, Howes, Tusting, Kang, Cameron, Bisanzio, Battle, Bhatt and
  Gething]{weiss2018global}
\textsc{Weiss, D~J, Nelson, A, Gibson, HS, Temperley, W, Peedell, S, Lieber, A,
  Hancher, M, Poyart, E, Belchior, S, Fullman, N, Mappin, B, Dalrymple, U,
  Rozier, J, Lucas, T C~D, Howes, R~E, Tusting, L~S, Kang, S~Y, Cameron, E,
  Bisanzio, D, Battle, K~E, Bhatt, S} \emph{and others}. (2018).
\newblock A global map of travel time to cities to assess inequalities in
  accessibility in 2015.
\newblock {\em Nature\/}~\textbf{553}(7688), 333.

\bibitem[Wilson and Wakefield(2020)Wilson and Wakefield]{wilson2020pointless}
\textsc{Wilson, Katie and Wakefield, Jon}. (2020).
\newblock Pointless spatial modeling.
\newblock {\em Biostatistics\/}~\textbf{21}(2), e17--e32.

\bibitem[{World Health Organization}(2019){World Health
  Organization}]{world2019world}
\textsc{{World Health Organization}}. (2019).
\newblock World malaria report 2019.

\end{thebibliography}
\FloatBarrier

\clearpage

\FloatBarrier

\markboth{Supplementary material}{Supplementary material}
\renewcommand{\thefigure}{S\arabic{figure}}

\FloatBarrier
\begin{figure}
\centering
   \includegraphics[width=0.9\linewidth]{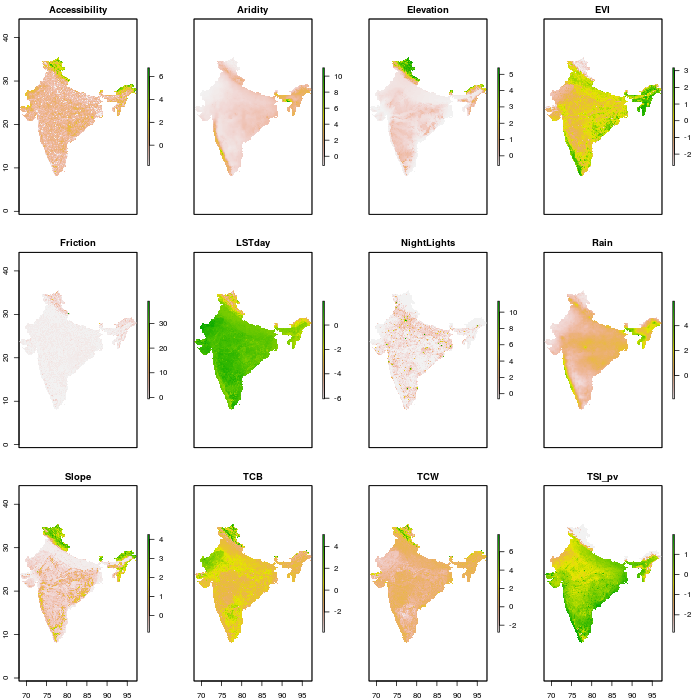}
   \caption{Real covariates used for India simulations.}
   \label{fig:covs_India} 
\end{figure}

\begin{figure}
\centering
   \includegraphics[width=0.9\linewidth]{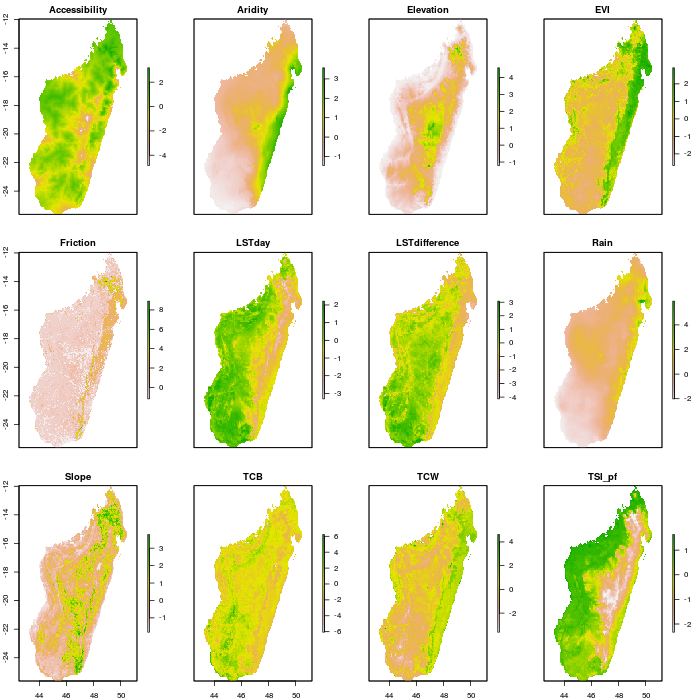}
   \caption{Real covariates used for Madagascar simulations.}
   \label{fig:covs_Mdg} 
\end{figure}

\begin{figure}
\centering
   \includegraphics[width=0.9\linewidth]{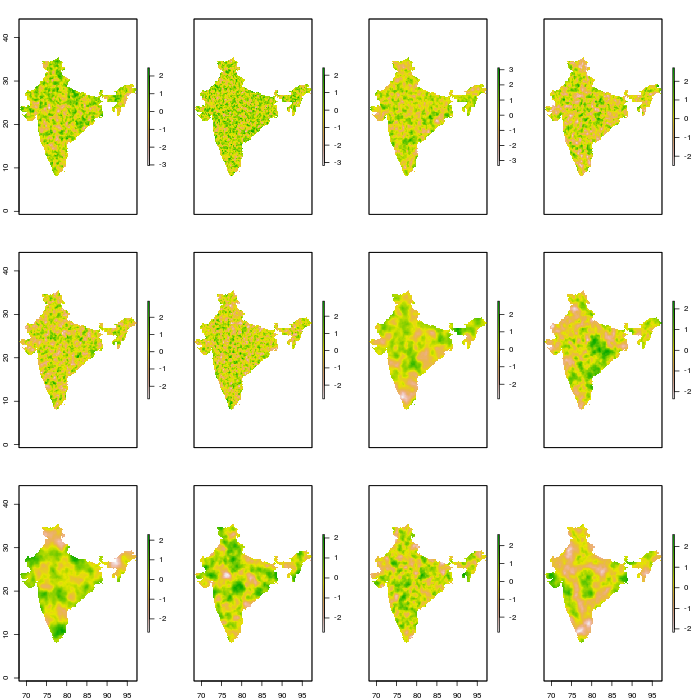}
   \caption{Mock covariates used for India simulations.}
   \label{fig:mock_covs_India} 
\end{figure}

\begin{figure}
\centering
   \includegraphics[width=0.9\linewidth]{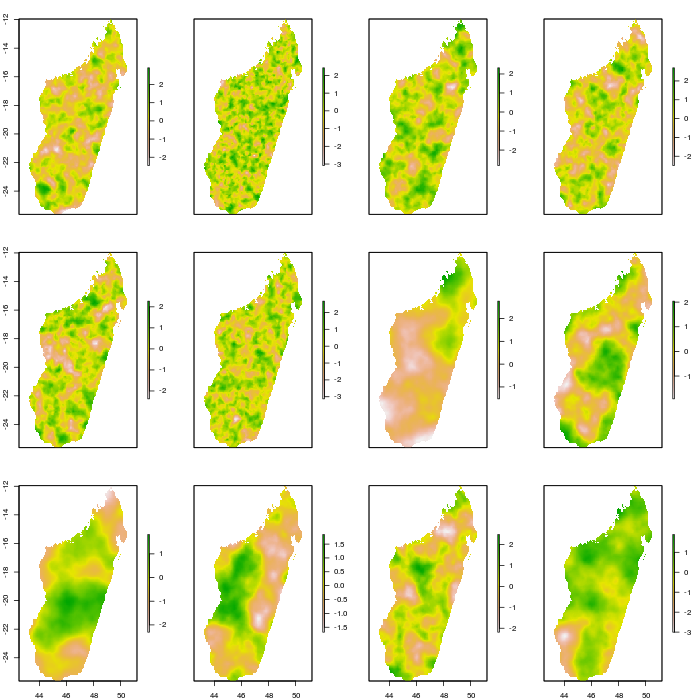}
   \caption{Mock covariates used for Madagascar simulations.}
   \label{fig:mock_covs_Mdg} 
\end{figure}

\end{document}